\documentclass[english,aps,manuscript]{revtex4}
\usepackage[T1]{fontenc}
\usepackage[latin9]{inputenc}
\usepackage{verbatim}
\usepackage{textcomp}
\usepackage{amssymb}
\usepackage{graphicx}

\makeatletter
\@ifundefined{textcolor}{}
{%
 \definecolor{BLACK}{gray}{0}
 \definecolor{WHITE}{gray}{1}
 \definecolor{RED}{rgb}{1,0,0}
 \definecolor{GREEN}{rgb}{0,1,0}
 \definecolor{BLUE}{rgb}{0,0,1}
 \definecolor{CYAN}{cmyk}{1,0,0,0}
 \definecolor{MAGENTA}{cmyk}{0,1,0,0}
 \definecolor{YELLOW}{cmyk}{0,0,1,0}
 }

\makeatother

\usepackage{babel}
\begin{document}

\title{PT Symmetry in Classical and Quantum Statistical Mechanics}

\author{Peter N. Meisinger and Michael C. Ogilvie}

\affiliation{Dept. of Physics, Washington University, St. Louis, MO 63130 USA}
\begin{abstract}
$\mathcal{PT}$-symmetric Hamiltonians and transfer matrices arise
naturally in statistical mechanics. These classical and quantum models
often require the use of complex or negative weights and thus fall
outside of the conventional equilibrium statistical mechanics of Hermitian
systems. $\mathcal{PT}$-symmetric models form a natural class where
the partition function is necessarily real, but not necessarily positive.
The correlation functions of these models display a much richer set
of behaviors than Hermitian systems, displaying sinusoidally-modulated
exponential decay, as in a dense fluid, or even sinusoidal modulation
without decay. Classical spin models with $\mathcal{PT}$ symmetry
include $Z(N)$ models with a complex magnetic field, the chiral Potts
model and the anisotropic next-nearest-neighbor Ising (ANNNI) model.
Quantum many-body problems with a non-zero chemical potential have
a natural $\mathcal{PT}$-symmetric representation related to the
sign problem. Two-dimensional QCD with heavy quarks at non-zero chemical
potential can be solved by diagonalizing an appropriate $\mathcal{PT}$-symmetric
Hamiltonian.
\end{abstract}


\maketitle

\section{Introduction}

The fundamental importance of $\mathcal{PT}$ symmetry was first pointed
out by Bender and Boettcher in their seminal work on quantum-mechanical
models \cite{Bender:1998ke}. Their work grew out of the observation
that the Hamiltonian 
\begin{equation}
H=p^{2}+igx^{3}.
\end{equation}
has only real eigenvalues. Bender and Boettcher observed that the
Hamiltonian $H$, while not Hermitian, is invariant under the simultaneous
application of the symmetry operations parity $\mathcal{P}:\, x\rightarrow-x$
and time reversal $\mathcal{T}:\, i\rightarrow-i$. This symmetry
ensures that all eigenvalues of $H$ are either real or part of a
complex pair. The argument is simple: if $H\left|\psi\right\rangle =E\left|\psi\right\rangle $
then $H\mathcal{PT\,\left|\psi\right\rangle =}\mathcal{PT}\, H\left|\psi\right\rangle =\mathcal{PT}\, E\left|\psi\right\rangle =E^{*}\mathcal{PT}\,\left|\psi\right\rangle $.
Thus if $E$ is an eigenvalue, $E^{*}$ is an eigenvalue as well.
The unexpected feature of the $ix^{3}$ model discovered by Bender
and Boettcher is that all of its eigenvalue are real. Since that discovery,
research on $\mathcal{PT}$-symmetric systems has grown enormously,
with applications in many areas of physics \cite{Bender:2005tb,Bender:2007nj}.

The $ix^{3}$ model has its origins in the study of the Lee-Yang theory
of phase transtions \cite{Yang:1952be,Lee:1952ig}. Lee and Yang showed
how the critical properties of the Ising model could be studied using
an analytic continuation of the external magnetic field $h$ to imaginary
values. They demonstrated that the zeros of the partition function
were all on the line $Re\left(h\right)=0$, reaching the real axis
at the critical point. It was subsequently realized that the phase
transition the edge of the gap in the distribution of zeros above
the critical temperature is itself a critical point described by an
$i\phi^{3}$ field theory \cite{Fisher:1978pf}. In two dimensions,
this field theory at its critical point yields the simplest of the
non-unitary minimal conformal field theories \cite{Cardy:1985yy,Cardy:1989fw,Zamolodchikov:1989cf,Yurov:1989yu}.
In one dimenson, the $i\phi^{3}$ field theory reduces to $ix^{3}$
quantum mechanics. There is a connection between $\mathcal{PT}$-symmetric
quantum mechanical models and  conformal field theories
in two dimensions \cite{Dorey:2009xa}. This connection is closely
related to the proof that a large class of $\mathcal{PT}$-symmetric
quantum mechanical models, including the $ix^{3}$ model, have only
real spectra \cite{Dorey:2001uw}; see also \cite{Dorey:2007zx}. 

Our focus here will be the application of generalized $\mathcal{PT}$
symmetry to some well-known models of statistical mechanics. By generalized
$\mathcal{PT}$ symmetry, we mean that the relevant unitary operator
is not necessarily the parity operator; in many cases, the role of
$\mathcal{P}$ will be played by the charge conjugation operator.
Some models, such as the $Z(N)$ spin model with a complex magnetic
field, will have a direct connection to Lee-Yang theory. In other
cases, such as the anisotropic next-nearest-neighbor Ising model (ANNNI
model) \cite{Selke:1988ss} or the chiral Potts model \cite{Ostlund:1981zz,Howes:1983mk},
no complex numbers will appear in the formulation of the model and
the appearance of $\mathcal{PT}$ symmetry will be hidden.

% added rev B
For those already familiar with the subject of ${\mathcal PT}$ symmetry, 
we should emphasize that the
focus here is somewhat different from the majority of work on the subject. 
We are interested in extant models in statistical mechanics 
in which PT symmetry plays a role, 
and not on the statistical mechanics of ${\mathcal PT}$-symmetric models {\it per se}.
There are some differences between the application of PT symmetry
to statistical mechanics and work on ${\mathcal PT}$-symmetric quantum mechanics.
For the models we discuss, the question of integration contours
in the complex plane does not arise.
Furthermore, the standard inner product,
as opposed to the CPT inner product often used in
${\mathcal PT}$-symmetric quantum mechanics,
 is sufficient.
Thus the models we discuss are similar
to the Ising model in an imaginary field,
which is the prototypical example of
${\mathcal PT}$ symmetry in statistical mechanics.

Our original interest in this subject was motivated by the sign problem,
particularly its appearance in QCD at finite density \cite{Lombardo:2008sc,Stephanov:2007fk}.
Essentially, the sign problem is really an instance of the general
problem of complex weights in a statistical sum. Such weights may
arise because the Hamiltonian or action is complex; as we explain
below, the problem arises naturally in Euclidean quantum field theories
with non-zero chemical potential. The sign problem is a large barrier
to first-principles lattice simulations of QCD at finite density and
the study of color superconductivity \cite{Alford:2007xm}. As we
show in section \ref{sec:Quantum-field-theory}, all Euclidean quantum
field theories with a sign problem due to a non-zero chemical potential
have a generalized $\mathcal{PT}$ symmetry. The class of statistical
models with generalized $\mathcal{PT}$ symmetry is very large. We
will show below that this class is precisely the set of models for
which the complex weight problem can be reduced to a sign problem,
\emph{i.e.}, there is a representation in which the partition function
is constructed from only real, but not necessarily positive, weights.
We believe that generalized $\mathcal{PT}$ symmetry is likely to
play a role in any future solution of the sign problem.

The spectral properties of statistical models with $\mathcal{PT}$
symmetry are different from those of Hermitian theories, which have
the property of spectral positivity. Essentially, this property ensures
that connected two-point correlation functions of observable will
fall monotonically to zero with separation. When there is a mass gap
between the ground state and the lowest excited state, this mass gap
determines a minimum rate of exponential decay for correlation functions.
Systems with generalized $\mathcal{PT}$ symmetry need not have spectral
positivity, and their two-point correlation functions can show sinusoidally-modulated
exponential decay, as in a dense fluid, or even sinusoidal modulation
without decay, as in periodic phases. Because these behaviors are
observed in nature, it is perhaps unsurprising that $PT$ symmetry
occurs in many problems of statistical mechanics. 

The remainder of this article is structured as follows: section II
introduces the fundamentals of generalized $\mathcal{PT}$ symmetry
and its role in the sign problem. Section III discusses the general
spectral properties of $\mathcal{PT}$-symmetric models. In section
IV, we treat several models from classical statistical mechanics.
Section V consider quantum statistical models. We show how $\mathcal{PT}$
symmetry makes it possible to solve a form of two-dimensional QCD
at finite density. A final section provides brief concluding remarks.

\section{\label{sec:Fundamentals}Fundamentals of $\mathcal{PT}$ symmetry
and the sign problem}

In this section, we discuss some fundamental aspects of $\mathcal{PT}$
symmetry. We also discuss here some fundamental aspects of the sign
problem that have a close relation with $\mathcal{PT}$ symmetry.
Although we will for the most part consider $\mathcal{PT}$-symmetric
Hamiltonians, in classical statistical mechanics it is often most
natural to work with $\mathcal{PT}$-symmetric transfer matrices,
and the results for Hamiltonians have obvious counterparts for transfer
matrices. Typically, the transfer matrix $T$ of a lattice model acts
as a Euclidean propagator similar to $\exp\left(-tH\right)$ in a
particular lattice direction. If the lattice has length $L$ in that
direction, then with periodic boundary conditions the partition function
is $Z=Tr\left(T^{L}\right)$. It is sometimes convenient to identify
$L$ with $\beta$ and $T$ with $\exp\left(-H\right)$ so that $Tr\left(T^{L}\right)$
can be written as $Tr\left(\exp\left(-\beta H\right)\right)$.

\subsection{Eigenvalues of $\mathcal{PT}$-symmetric systems}

Given a Hilbert space, the adjoint $H^{+}$ of an operator $H$ is
defined using the inner product:
\begin{equation}
\left\langle \phi\left|H^{+}\right|\psi\right\rangle =\left\langle \psi\left|H\right|\phi\right\rangle ^{*}
\end{equation}
 which we often write as
\begin{equation}
H^{+}=H^{*T}.
\end{equation}
Operators satisfying $H^{+}=H$ are said to be Hermitian (we follow
the usual physics practice and do not distinguish between Hermitian
and self-adjoint operators). It is a standard result that the eigenvalues
of a Hermitian operator are real. Operators with a generalized $\mathcal{PT}$
symmetry have a more general constraint on their eigenvalues: they
are either real or form a complex conjugate pair.\emph{ }%
\begin{comment}
\emph{From 2009 PT Theory ms:}
\end{comment}
{} A Hamiltonian $H$ is $\mathcal{PT}$-symmetric if $\left[H,\mathcal{PT}\right]=0$.
Let $\left|\psi\right\rangle $ be an eigenstate of $H$ with eigenvalue
$E$. Then we have $ $$H\mathcal{PT\,\left|\psi\right\rangle =}\mathcal{PT}\, H\left|\psi\right\rangle =\mathcal{PT}\, E\left|\psi\right\rangle =E^{*}\mathcal{PT}\,\left|\psi\right\rangle $.
Thus we see that $\mathcal{PT}\,\left|\psi\right\rangle $ is an eigenstate
of $H$ with eigenvalue $E^{*}$. If $\left|\psi\right\rangle $ is
an eigenstate of $\mathcal{PT}$ , then necessarily $E$ is real.
However, $\mathcal{PT}$- symmetric Hamiltonians can also have complex
conjugate pairs of eigenvalues. The case where two or more eigenvalues
are not real is usually described as broken $\mathcal{PT}$ symmetry.
Eigenstates of $H$ associated with a complex eigenvalue pair cannot
be eigenstates of $\mathcal{PT}$. In the models we consider, $\mathcal{T}$
is implemented as complex conjugation, and $\mathcal{P}$ is a unitary
operator obeying $\left[\mathcal{P},\mathcal{T}\right]=0$ and $\mathcal{P}^{2}=1$.

\subsection{Bender-Mannheim theorem}

Not all ${\mathcal PT}$-symmetric models are obviously so. A simple criterion for
a $\mathcal{PT}$-symmetric Hamiltonian $H$ (or transfer matrix $T$)
has been given by Bender and Mannheim \cite{Bender:2009mq}. If the
characteristic polynomial $\det\left[H-\lambda I\right]$ has real
coefficients, then $H$ has a generalized $\mathcal{PT}$ symmetry.
Interesting models arise in statistical mechanics with hidden $\mathcal{PT}$
symmetry when the transfer matrix $T$ is real but not symmetric,
as will be discussed below in section \ref{sec:Classical-statistical-mechanics}.

The striking feature of the $ix^{3}$ models is that all of its energy
eigenvalues are real. This is usually referred to as unbroken $\mathcal{PT}$
symmetry. We will distinguish three different behaviors of $\mathcal{PT}$-symmetric
models, based on the reality of the eigenvalues of the Hamiltonian
$H$ or transfer matrix $T$. We order the eigenvalues of $H$ by their
real parts, so that the ground state of $H$ has the eigenvalue with
the lowest real part. Similarly, we order the eigenvalues of $T$ by
their magnitude; the eigenvalue largest in magnitude is the analog
of the ground state energy.
Typically the different behaviors are
each associated with a different part of parameter space. 
In region
I, $\mathcal{PT}$ symmetry is unbroken, and all eigenvalues are real.
In region II, the lowest eigenvalue of $H$ is real, but $\mathcal{PT}$
symmetry is broken by one or more pairs of excited states becoming
complex. For a transfer matrix $T$, the eigenvalue largest in absolute
value is real in region II, but other eigenvalues are complex. In
region III, $\mathcal{PT}$ symmetry is broken by the ground state
of $H$ becoming complex. For a transfer matrix $T$, the eigenvalue %Fixed eigenvale
of largest absolute value becomes complex. Strictly speaking,
by our definition above this means there are two ground states.
This does not occur in conventional quantum-mechanical systems
with a finite number of degrees of freedom, but does occur in
$\mathcal{PT}$-symmetric models. It also occurs, of course,
in Hermitian models with an infinite number of degrees of
freedom, and is the basis of spontaneous symmetry breaking.
As we show in the next
section, this distinction is physical, and manifests directly in correlation
functions. In region II, oscillatory behaviors appears in correlation
functions. In region III, the system is in a spatially modulated phase.
We emphasize that the behavior of correlation functions seen in regions
II and III cannot be obtained from conventional models for which $H$
is Hermitian: such behavior is incompatible with the spectral representation
of the correlation function for Hermitian theories.

\subsection{Connection to complex weight problem}

A naive treatment of $\mathcal{PT}$-symmetric models often involves
a sum over complex weights, as in the path integral treatment of the
$ix^{3}$ model or in the continuation of the Ising model to imaginary
magnetic field. In many areas of physics, there are problems where
we wish to evaluate the {}``average'' of some quantity $x$ over
an ensemble with complex weights $w_{j}\in\mathbf{\mathbb{C}}$ :
\begin{equation}
\left\langle x\right\rangle \equiv\frac{\sum_{j}x_{j}w_{j}}{\sum_{j}w_{j}}
\end{equation}
There is no effective general algorithm for calculating such sums,
as is the case for positive weights; see, \emph{e.g.}, \cite{Loh:1990zz,Cox:1999nt,Chandrasekharan:2008gp}.
This problem is generally referred to as the sign problem, because
even the case of negative weights is difficult. We will discuss the
sign problem in detail for several models in sections \ref{sec:Classical-statistical-mechanics}
and \ref{sec:Quantum-field-theory}, but here discuss the general
role of $\mathcal{PT}$symmetry. Let us consider the case where the
$w_{j}$'s can be written as $\exp\left(-\beta E_{j}\right)$, as
is typical in statistical mechanics where $E_{j}$ is eigenvalue of
some operator $H$ and $\beta$ is the inverse of the temperature
$T$. For a Hermitian system, the partition function
\begin{equation}
Z\left(\beta\right)=\sum_{j}e^{-\beta E_{j}}
\end{equation}
 is real and positive for all real values of $\beta$, because the
eigenvalues are all real. On the other hand, if $H$ is $\mathcal{PT}$-symmetric,
the eigenvalues are either real or occur in complex conjugate pairs.
We will prove in sections \ref{sec:Spectral-properties-of} that $Z$
may be represented as
\begin{equation}
Z\left(\beta\right)=\sum_{real}e^{-\beta E_{j}}+\sum_{pairs}\left(e^{-\beta E_{k}}+e^{-\beta E_{j}^{*}}\right)
\end{equation}
which is real but not necessarily positive. As a consequence of the
Bender-Mannheim theorem, we have the following characterization of
partition functions: Models which are Hermitian, or equivalent to
Hermitian models under a similarity transform, have $Z$ real and
positive for all real $\beta$; models which have generalized $\mathcal{PT}$
symmetry have $Z$ real, but not necessarily positive, for all real
$\beta$. A Laplace transform argument shows that the converse is
also true. Thus models with generalized $\mathcal{PT}$ symmetry are
precisely the class of models in which the complex weight problem
can be reduced a genuine sign problem, the problem of averaging over
positive and negative weights.

\subsection{Equivalence to Hermitian if $\mathcal{PT}$ symmetry is unbroken}

\emph{}%
\begin{comment}
Towards a solution of the sign problem \emph{From PT Theory Letter
2010}
\end{comment}

The difficulty presented by the sign problem depends directly on $\mathcal{PT}$
symmetry breaking or its absence. Mostafazadeh \cite{Mostafazadeh:2003gz}
has proven that when $\mathcal{PT}$ symmetry is unbroken (region
I) and the spectrum is non-degenerate,
there is a similarity transformation $S$ that transforms a $\mathcal{PT}$-symmetric
Hamiltonian $H$ into an isospectral Hermitian Hamiltonian $H_{h}$
via $H_{h}=SHS^{-1}$. If $S$ can be found, the transformation of
$H$ into $H_{h}$ eliminates the sign problem for $\mathcal{PT}$-symmetric
quantum Hamiltonians throughout the interior of region I.
On the boundary of region I, where two or more
eigenvalues become degenerate, it is possible
that the Hamiltonian can be of a non-diagonalizable
Jordan block form \cite{PhysRevD.78.025022}.
The equivalence to a Hermitian Hamiltonian also applies
to $\mathcal{PT}$-symmetric transfer matrices $T$, but a further
restriction to positive eigenvalues for $T$ is necessary for the
elimination of the sign problem. Thus there are regions of parameter
space where the sign problem can be removed by a similarity transformation.
Unfortunately, the explicit construction of the similarity transform
typically requires knowledge of the exact eigenvalues and eigenvectors.
However, there are some models, notably the $-\lambda x^{4}$ model,
for which the similarity transform or an equivalent functional integral
transformation is known \cite{Andrianov:2007vt,Bender:2006wt,Jones:2006et,Ogilvie:2008tu}.
In regions II and III, the sign problem has an underlying physical
basis, and cannot be removed by a similarity transformation. The negative
weight contributions to the partition function $Z$ arise from the
contributions of complex conjugate eigenvalue pairs associated with
$\mathcal{PT}$ symmetry breaking. It is that breaking that in turn
gives rise to the oscillatory and damped oscillatory behavior of two-point
functions characteristic of many physical systems.

\subsection{Real Representations of $\mathcal{PT}$-symmetric Hamiltonians}

There are $\mathcal{PT}$ symmetric models like the ANNNI model \cite{Selke:1988ss}
where the classical Hamiltonian, corresponding to the action in the
path integral formalism, is real. Such a model can be simulated with
no difficulties of principle throughout its parameter space. In Hermitian
systems, the existence of an antiunitary involution commuting with
the Hamiltonian implies that there is a basis in which $H$ is real;
see, \emph{e.g.}, \cite{Haake:2010}. This theorem easily extends
to the case of those $\mathcal{PT}$-symmetric systems for which $\left(\mathcal{PT}\right)^{2}=1$,
and can be applied to transfer matrices as well as Hamiltonians. This
suggests the existence of a class of $\mathcal{PT}$-symmetric models
which can be simulated in all three regions, but no general criterion
for determining the class is known.

The proof that for any $\mathcal{PT}$-symmetric Hamiltonian there
is a basis in which the matrix elements of $H$ are real follows closely
the proof for Hermitian systems \cite{Haake:2010}. The antiunitary
operator $\mathcal{PT}$ commutes with the Hamiltonian $[\mathcal{PT},H]=0$
and satisfies $\left(\mathcal{PT}\right)^{2}=1$. These properties
are sufficient to construct a real representation of the Hamiltonian
$H$. The first step is the construction of a $\mathcal{PT}$-invariant
basis $\psi_{a}$. We start from any non-zero vector $\phi_{1}$ and
complex number $\alpha_{1}.$ The vector $\psi_{1}=\alpha_{1}\phi_{1}+\mathcal{PT}\alpha_{1}\phi_{1}$
is invariant under $\mathcal{PT}$. Choosing a vector $\phi_{2}$
orthogonal to $\psi_{1}$, we form the vector $\psi_{2}=\alpha_{2}\phi_{2}+\mathcal{PT}\alpha_{2}\phi_{2}$,
where $\alpha_{2}$. The inner product of $\psi_{2}$ and $\psi_{1}$
is zero:
\begin{equation}
\left\langle \psi_{2}|\psi_{1}\right\rangle =\alpha_{2}^{*}\left\langle \phi_{2}|\psi_{1}\right\rangle +\alpha_{2}\left\langle \mathcal{PT}\phi_{2}|\psi_{1}\right\rangle =\alpha_{2}^{*}\left\langle \left(\mathcal{PT}\right)^{2}\phi_{2}|\psi_{1}\right\rangle +\alpha_{2}\left\langle \mathcal{PT}\psi_{2}|\psi_{1}\right\rangle 
\end{equation}
\begin{eqnarray*}
\left\langle \psi_{2}|\psi_{1}\right\rangle  & = & \alpha_{2}^{*}\left\langle \phi_{2}|\psi_{1}\right\rangle +\alpha_{2}\left\langle \mathcal{PT}\phi_{2}|\psi_{1}\right\rangle \\
 & = & 0+\alpha_{2}\left\langle \left(\mathcal{PT}\right)^{2}\phi_{2}|\mathcal{PT}\psi_{1}\right\rangle ^{*}\\
 & = & \alpha_{2}\left\langle \phi_{2}|\psi_{1}\right\rangle ^{*}=0
\end{eqnarray*}
The complex number $\alpha_{2}$ can be adjusted to ensure that $\psi_{2}$
is not the zero vector. By proceeding in this fashion, a $\mathcal{PT}$-symmetric
orthogonal basis can be constructed. In this basis, the matrix elements
of the Hamiltonian are real:
\begin{eqnarray*}
H_{ab} & = & \left\langle \psi_{a}|H\psi_{b}\right\rangle \\
 & = & \left\langle \mathcal{PT}\psi_{a}|\mathcal{PT}H\psi_{b}\right\rangle ^{*}\\
 & = & \left\langle \mathcal{PT}\psi_{a}|H\left(\mathcal{PT}\right)\psi_{b}\right\rangle ^{*}\\
 & = & \left\langle \psi_{a}|H\psi_{b}\right\rangle ^{*}\\
 & = & H_{ab}^{*}
\end{eqnarray*}
Note that this results holds independent of the appearance of complex
eigenvalues in $H$. Because $H$ has only real matrix elements in
this basis, it is clear that that the secular $\det\left(z-H\right)=0$
of such a system has only real coefficients, a fact closely linked
to $\mathcal{PT}$ symmetry by the Bender-Mannheim theorem.

\begin{comment}
Fritz Haake, Quantum Signatures of Chaos, 2nd ed., p. 20
\end{comment}

\section{\label{sec:Spectral-properties-of}Spectral properties of ${\mathcal PT}$-symmetric
models}

Spectral positivity plays a fundamental role in Hermitian systems.
As we have seen, it ensures that the partition function $Z$ is always
positive, and it gives a representation of two-point functions as
sums of decaying exponentials. On a practical level, it allows for
the isolation of the lightest state in a given channel from the large-distance
behavior of two-point functions. Because $\mathcal{PT}$-symmetric
models lead naturally to two different basis sets, completeness and
the subsequent derivation of the Kallen-Lehmann representation for
two-point functions is more complicated than in the Hermitian case.
We give a self-contained derivation of both below, followed by a brief
discussion of the implications of our results. Of particular interest
is the connection of broken $\mathcal{PT}$ symmetry of the ground
state (region III), with generalized Yang-Lee phase transitions. For
simplicity, we will use $H$ and $\beta$ throughout this section,
but we note that in classical statistical mechanics problems, the
relevant objects are $T$ and $L$, in which case $\beta$ is not
the inverse temperature of the system and correlation functions are
between different spatial locations. As we will see for two-dimensional
QCD in section \ref{sec:Quantum-field-theory}, a transfer matrix
approach can also be useful for quantum systems as well.

\subsection{Completeness and Kallen-Lehman representation for $\mathcal{PT}$-symmetric
systems}

We now prove a completeness relation for $\mathcal{PT}$-symmetric
models which is valid in all three regions. We consider the typical
case where the Hamiltonian $H$ is diagonalizable via a similarity
transformation and has discrete, non-degenerate eigenvalues.
Exceptional points, in the sense of degenerate real eigenvalues, occur at the boundary of region I as well as inside region II and III.  There are well-known difficulties in the spectral resolution at these points; 
see \cite{Andrianov:2010an} and \cite{PhysRevD.78.025022} for specific examples as well as
the general discussion in \cite{Bender:2009mq}.
The passage from region I to region III, where the ground state eigenvalue becomes degenerate, is a critical point analogous to those found in Hermitian models; the best-known example is the Yang-Lee singularity. In analogy with symmetry-breaking behavior in Hermitian models, it is likely necessary to define the behavior at such a point using either an infinitesimal symmetry-breaking perturbation or symmetry-breaking boundary conditions. 
At this stage of our understanding of ${\mathcal PT}$-symmetric models in statistical mechanics, we cannot provide a general prescription, but must handle each model on a case-by-case basis.
The case of non-degenerate eigenvalues, in contrast, is tractable for all models.
By virtue
of the secular equation for the eigenvalues, $H$ and $H^{T}$ are
isospectral: the existence of an eigenvalue-eigenvector pair for $H$
\begin{equation}
H\left|j\right\rangle =E_{j}\left|j\right\rangle 
\end{equation}
implies the existence of a corresponding pair for $H^{T}$:
\begin{equation}
H^{T}\left|\tilde{j}\right\rangle =E_{j}\left|\tilde{j}\right\rangle .
\end{equation}
From the commutation relation $\left[H,\mathcal{PT}\right]=0$, we
have
\begin{equation}
H\mathcal{P}\left|\mathcal{T}j\right\rangle =H\mathcal{P}\mathcal{T}\left|j\right\rangle =\mathcal{P}\mathcal{T}H\left|j\right\rangle =\mathcal{P}\mathcal{T}E_{j}\left|j\right\rangle =E_{j}^{*}\mathcal{P}\left|\mathcal{T}j\right\rangle 
\end{equation}
 and its Hermitian conjugate
\begin{equation}
\left\langle \mathcal{T}j\right|\mathcal{P}H^{\dagger}=E_{j}\left\langle \mathcal{T}j\right|\mathcal{P}.
\end{equation}
Noting that $\mathcal{P}H\mathcal{P}=H^{*}$ we have
\begin{equation}
\left\langle \mathcal{T}j\right|\mathcal{P}H^{\dagger}=\left\langle \mathcal{T}j\right|\mathcal{P}\left(\mathcal{P}H\mathcal{P}\right)^{T}=\left\langle \mathcal{T}j\right|H^{T}\mathcal{P}
\end{equation}
 so that we find after multiplying by $\mathcal{P}$ on the right
that
\begin{equation}
\left\langle \mathcal{T}j\right|H^{T}=E_{j}\left\langle \mathcal{T}j\right|
\end{equation}
 Now consider the matrix element
\begin{equation}
\left\langle \mathcal{T}j\left|\left(H^{T}-H^{T}\right)\right|\tilde{k}\right\rangle =\left(E_{j}-E_{k}\right)\left\langle \mathcal{T}j|\tilde{k}\right\rangle =0
\end{equation}
 The set of eigenstates $\left\{ \left|j\right\rangle \right\} $
and $\left\{ \left|\tilde{k}\right\rangle \right\} $ are both complete
so we must have
\begin{equation}
\left\langle \mathcal{T}j|\tilde{k}\right\rangle \propto\delta_{jk}
\end{equation}
 and also
\begin{equation}
\left\langle \mathcal{T}\tilde{k}|j\right\rangle \propto\delta_{jk}
\end{equation}
 Thus we have two completeness relations
\begin{equation}
\sum_{j}\frac{\left|j\right\rangle \left\langle \mathcal{T}\,\tilde{j}\right|}{\left\langle \mathcal{T}\tilde{j}|j\right\rangle }=1
\end{equation}
 and 
\begin{equation}
\sum_{j}\frac{\left|\tilde{j}\right\rangle \left\langle \mathcal{T}\, j\right|}{\left\langle \mathcal{T}j|\tilde{j}\right\rangle }=1.
\end{equation}
% added rev B
These completeness relations assume the Hamiltonian H  acts in a space in which the inner products 
$\left\langle n|\mathcal{T}\tilde{n}\right\rangle$   are finite. The consequences of this assumption can vary from model to model. 
See \cite{Bender:1998ke} for a specific example where the contour along which a ${\mathcal PT}$-symmetric differential equation is evaluated is deformed into the complex plane to ensure finite inner products. However, this technical point will not be of concern for the models we analyze in this paper.
We will generally choose the normalization of states such that $\left\langle \mathcal{T}\tilde{j}|j\right\rangle =\left\langle \mathcal{T}j|\tilde{j}\right\rangle =1$. 

With completeness relations at hand we can now calculate the partition
function and spatial two-point functions in terms of eigenvalues and
eigenstates. For a quantum theory, the partition function $Z$ is
given by
\begin{equation}
Z=\sum_{n}\left\langle n\right|e^{-\beta H_{PT}}\left|n\right\rangle 
\end{equation}
where $\left\{ \left|n\right\rangle \right\} $ is an arbitrary orthonormal
basis; note that the basis formed by the eigenstates of $H_{PT}$
cannot be used in this way to define $Z$ 
unless $H_{PT}$ is Hermitian.
% rev B
When $H$ is not Hermitian and PT symmetry is unbroken,
it is a common practice to require
$\mathcal{PT}\left|j\right\rangle =\left|j\right\rangle$   and to introduce an additional linear operator $\mathcal{C}$  
such that $\mathcal{CPT}\left|j\right\rangle \cdot\left|k\right\rangle =\delta_{jk}$  \cite{Bender:2005tb,Bender:2007nj}. 
Using these conventions, one can write the partition function as a sum over eigenstates
of the form $Z=\sum_{j}\left\langle j\right|Ve^{-\beta H}\left| j\right\rangle$   
where $V=\left(\mathcal{CP}\right)^{T}$. 
The need for the operator $V$  arises because the eigenstates $\left|j\right\rangle$   
are not orthonormal with respect to the inner product $\left\langle m|n\right\rangle$. However, if $H$  breaks PT-symmetry, 
not all of the eigenstates of $H$  are eigenstates of $\mathcal{PT}$, and an alternative expression for $V$  of the form
$V=\sum_{j}\left|\mathcal{T}\tilde{j}\right\rangle \left\langle \mathcal{T}\tilde{j}\right|$ would need to be used. We avoid these considerations in this paper by performing the trace in the partition function over an arbitrary orthonormal basis; 
such a basis may be  constructed from the eigenstates $\left| j\right\rangle$ using the Gram-Schmidt algorithm.

We insert the $\mathcal{PT}$ completeness relation
\begin{equation}
Z=\sum_{n}\left\langle n\right|e^{-\mathbf{\beta}H}\sum_{j}\frac{\left|j\right\rangle \left\langle \mathcal{T}\,\tilde{j}\right|}{\left\langle \mathcal{T}\tilde{j}|j\right\rangle }\left|n\right\rangle =\sum_{j}e^{-\beta E_{j}}\frac{\sum_{n}\left\langle n|j\right\rangle \left\langle \mathcal{T}\,\tilde{j}|n\right\rangle }{\left\langle \mathcal{T}\tilde{j}|j\right\rangle }=\sum_{j}e^{-\beta E_{j}}
\end{equation}
so that $Z$ has the same form as in Hermitian theories, as assumed
in section \ref{sec:Fundamentals}. We can write $Z$ usefully as
\begin{equation}
Z=\sum_{r}e^{-\beta E_{r}}+\sum_{p}\left(e^{-\beta E_{p}}+e^{-\beta E_{p}^{*}}\right)
\end{equation}
where the sum over $r$ is over all real energies $E_{r}$ and the
sum over $p$ is over pairs of complex energies. The oscillatory character
of the second sum leads to negative contributions to the partition
function, which is the sign problem. Note that in region I, there
is no sign problem. The sign problem only arises in region II and
III. Strictly speaking, the sign problem disappears in region II in
the limit $\beta\rightarrow\infty$. For transfer matrix problems,
this implies that in the limit $L\rightarrow\infty$, the sign problem
becomes negligible in region II. Only in region III does the sign
problem survive the infinite volume limit in the calculation of $Z$.

\subsection{Spectral theorems}

We now prove some general results for the typical case where $H$
is symmetric but complex. When $H$ is symmetric, we have $\left|j\right\rangle =\left|\tilde{j}\right\rangle $
and $\left|\mathcal{T}j\right\rangle =\left|\mathcal{T}\tilde{j}\right\rangle $
and we immediately obtain the somewhat simpler completeness relation
\begin{equation}
\sum_{j}\left|j\right\rangle \left\langle \mathcal{T}\, j\right|=1
\end{equation}
In typical $\mathcal{PT}$-symmetric quantum mechanics models, $H$
is symmetric, and the completeness relation takes the form \cite{Bender:2005tb,Bender:2007nj}
\begin{equation}
\sum_{j}\frac{\psi_{j}\left(x\right)\psi_{j}\left(y\right)}{\int dx\,\psi_{j}^{2}\left(x\right)}=\delta\left(x-y\right)
\end{equation}
showing the explicit normalization factor.

We begin by proving a result for certain matrix elements that appear
repeatedly in these calculations. If $E_{j}\neq E_{j}^{*}$, then
the properly normalized eigenstate of $H$ with eigenvalue $E_{j}^{*}$
is related to $\mathcal{PT}\left|j\right\rangle $ by a phase factor:

\begin{equation}
\mathcal{PT}\left|j\right\rangle =e^{i\alpha_{j}}\left|j^{*}\right\rangle .
\end{equation}
It immediately follows that

\begin{equation}
\mathcal{PT}\left|j^{*}\right\rangle =\mathcal{PT}e^{-i\alpha_{j}}\mathcal{PT}\left|j\right\rangle =e^{i\alpha_{j}}\left|j\right\rangle ,
\end{equation}
or

\begin{equation}
e^{i\alpha_{j^{*}}}=e^{i\alpha_{j}}.
\end{equation}
 On the other hand, if $E_{j}$ is real, we have simply
\begin{equation}
\mathcal{PT}\left|j\right\rangle =e^{i\alpha_{j}}\left|j\right\rangle .
\end{equation}
so in this case we may identify $\left|j^{*}\right\rangle $ with
$\left|j\right\rangle $. We will prove results for matrix elements
in the case where $E_{j}$ is complex, and similar results will hold
when $E_{j}$ is real with $\left|j^{*}\right\rangle $ replaced by
$\left|j\right\rangle $.

Consider the action of the bra

\begin{equation}
\left\langle \mathcal{T}j^{*}\right|\mathcal{PT}
\end{equation}
on an arbitrary ket

\begin{equation}
\left|\psi\right\rangle =\sum_{pairs}\left(\chi_{k}\left|k_{c}\right\rangle +\chi_{k^{*}}\left|k_{c}^{*}\right\rangle \right)+\sum_{reals}\rho_{n}\left|n_{r}\right\rangle .
\end{equation}
We have

\begin{equation}
\left\langle \mathcal{T}j^{*}\right|\mathcal{PT}\left|\psi\right\rangle =\left\langle \mathcal{T}j^{*}\right|\mathcal{PT}\left[\sum_{pairs}\left(\chi_{k}\left|k_{c}\right\rangle +\chi_{k^{*}}\left|k_{c}^{*}\right\rangle \right)+\sum_{reals}\rho_{n}\left|n_{r}\right\rangle \right],
\end{equation}

\begin{equation}
\left\langle \mathcal{T}j^{*}\right|\mathcal{PT}\left|\psi\right\rangle =\left\langle \mathcal{T}j^{*}\right|\left[\sum_{pairs}\left(\chi_{k}^{*}e^{i\alpha_{k}}\left|k_{c}^{*}\right\rangle +\chi_{k^{*}}^{*}e^{i\alpha_{k}}\left|k_{c}\right\rangle \right)+\sum_{reals}\rho_{n}^{*}e^{i\alpha_{n}}\left|n_{r}\right\rangle \right],
\end{equation}

\begin{equation}
\left\langle \mathcal{T}j^{*}\right|\mathcal{PT}\left|\psi\right\rangle =e^{i\alpha_{j}}\chi_{j}^{*},
\end{equation}
and

\begin{equation}
\left\langle \mathcal{T}j^{*}\right|\mathcal{PT}\left|\psi\right\rangle =e^{i\alpha_{j}}\left\langle Tj|\mathcal{\psi}\right\rangle ^{*}.
\end{equation}
 Now we apply the above result to a state of the form $\phi\left|j^{*}\right\rangle $
where $\mathcal{PT}\phi\mathcal{PT}=\phi$: 
\begin{equation}
\left\langle \mathcal{T}k^{*}\left|\phi_{1}\right|j^{*}\right\rangle =\left\langle \mathcal{T}k^{*}\left|PT\phi_{1}PT\right|j^{*}\right\rangle 
\end{equation}
\begin{equation}
\left\langle \mathcal{T}k^{*}\left|\phi_{1}\right|j^{*}\right\rangle =\left\langle \mathcal{T}k^{*}\left|PT\phi_{1}e^{i\alpha_{j}}\right|j\right\rangle 
\end{equation}
\begin{equation}
\left\langle \mathcal{T}k^{*}\left|\phi_{1}\right|j^{*}\right\rangle =e^{i\alpha_{j}}\left\langle \mathcal{T}k\left|\phi_{1}e^{i\alpha_{j}}\right|j\right\rangle ^{*}
\end{equation}
\begin{equation}
\left\langle \mathcal{T}k^{*}\left|\phi_{1}\right|j^{*}\right\rangle =\left\langle \mathcal{T}k\left|\phi_{1}\right|j\right\rangle ^{*}
\end{equation}
This result has been proven for the case where both $\left|j\right\rangle $
and $\mbox{\ensuremath{\left|k\right\rangle }}$ are half of a conjugate
pair of eigenvectors. It also holds when one or both of the eigenvectors
has a real eigenvalue, in which case we identify $\left|j^{*}\right\rangle $
with $\left|j\right\rangle $. In the case where both $E_{j}$ and
$E_{k}$ are real, the matrix element is real:
\begin{equation}
\left\langle \mathcal{T}k\left|\phi_{1}\right|j\right\rangle =\left\langle \mathcal{T}k\left|\phi_{1}\right|j\right\rangle ^{*}.
\end{equation}

%For
%simplicity, we will use $H$ and $\beta$ throughout this section,
%but we note that in classical statistical mechanics problems, the
%relevant objects are $T$ and $L$, in which case $\beta$ is not
%the inverse temperature of the system and correlations funtions are
%between different spatial locations. As we will see for two-dimensional
%QCD in section \ref{sec:Quantum-field-theory}, a transfer matrix
%approach can also be useful for quantum systems as well.

With these results in hand, we now turn to one- and two-point functions
in the case that $H$ is symmetric and fields satisfy $\mathcal{PT}\phi\mathcal{PT}=\phi$.
We remind the reader again that in application to classical statistical mechanics,
the Hamiltonian is related to the transfer matrix by $T=\exp\left(-H \right)$, and $\beta$
is identified with the length $L$ of the system. 
A one-point function can be written as
\begin{eqnarray}
\left\langle \phi\left(x\right)\right\rangle  & = & \frac{1}{Z}Tr\left[e^{-rH_{PT}}\phi e^{-\left(\beta-r\right)H_{PT}}\right]\nonumber\\
 & = & \frac{1}{Z}\sum_{jk}\left\langle \mathcal{T}\, j\right|e^{-rH_{PT}}\left|k\right\rangle \left\langle \mathcal{T}\, k\right|\phi e^{-\left(\beta-r\right)H_{PT}}\left|j\right\rangle \nonumber\\
 & = & \frac{1}{Z}\sum_{jk}\left\langle \mathcal{T}\, j\right|e^{-rE_{k}}\left|k\right\rangle \left\langle \mathcal{T}\, k\right|\phi e^{-\left(\beta-r\right)E_{j}}\left|j\right\rangle \nonumber\\
 & = & \frac{1}{Z}\sum_{j}e^{-\beta E_{j}}\left\langle \mathcal{T}\, j\right|\phi\left|j\right\rangle \nonumber\\
 & = & \frac{1}{Z}\left[\sum_{pairs}\left(e^{-\beta E_{j}}\left\langle \mathcal{T}\, j\right|\phi\left|j\right\rangle +e^{-\beta E_{j}^{*}}\left\langle \mathcal{T}\, j^{*}\right|\phi\left|j^{*}\right\rangle \right)+\sum_{reals}e^{-\beta E_{j}}\left\langle \mathcal{T}\, j\right|\phi\left|j\right\rangle \right]
\end{eqnarray}
where we have used $r$ to label the coordinate of $x$ in the direction
along which $H$ acts.
This expression for $\left\langle \phi\left(x\right)\right\rangle$  is manifestly real as a consequence of our results for matrix
elements.

Similarly, suppose $x$ and $y$ are two points separated by a distance $r$
in the direction in which $H$ acts. Then the two point function
for operator $\phi_{1}\left(x\right)$ and $\phi_{2}\left(y\right)$is
defined by
\begin{eqnarray}
\left\langle \phi_{1}\left(x\right)\phi_{2}\left(y\right)\right\rangle  & = & \frac{1}{Z}Tr\left[\phi_{1}e^{-rH_{PT}}\phi_{2}e^{-\left(\beta-r\right)H_{PT}}\right]\nonumber\\
 & = & \frac{1}{Z}\sum_{jk}e^{-rE_{j}}e^{-\left(\beta-r\right)E_{k}}\left\langle \mathcal{T}k\left|\phi_{1}\right|j\right\rangle \left\langle \mathcal{T}j\left|\phi_{2}\right|k\right\rangle 
\end{eqnarray}
 which is real using the same reasoning that was applied to the one-point
function. This representation establishes the fundamental observable
distinction between regions I, II and III. Each term in the spectral
representation of the two-point function depends on $r$ as $\exp\left[\left(E_{k}-E_{j}\right)r\right]$.
In regions I and II, the ground state is unique, and the terms with
$k=0$ dominate for large $\beta$. In region I, this leads to monotonic
exponential decay. In region II, some of the excited states have complex
energies, leading to modulated exponential decay in two-point functions.

In region III, the ground state, defined as the state with the lowest
value of $Re\left(E\right)$, is no longer unique. The states $\left|0\right\rangle $
and $\left|0^{*}\right\rangle $ will dominate in both $Z$ and in
two-point functions in the limit $\beta\rightarrow\infty$, or in
the limit $L\rightarrow\infty$ for transfer matrices. We can take
$\beta$ sufficiently large that all states except $E_{0}$ and $E_{0}^{*}$
can be neglected, in which case $Z$ can be approximated by
\begin{equation}
Z\simeq e^{-\beta E_{0}}+e^{-\beta E_{0}^{*}}
\end{equation}
and the approximate zeros of the partition function will occur at
\begin{equation}
Im\left(E_{0}\right)=\frac{\left(2p+1\right)\pi}{2\beta}
\end{equation}
where $p$ is any integer. This is consistent with a general theory
of partition function zeros that can be applied to models with $\mathcal{PT}$-symmetric
transfer matrices \cite{Biskup:2000bb}. Under some technical conditions,
the partition function in a periodic volume $V=L^{d}$ can be written
as
\begin{equation}
Z=\sum_{m}e^{-\beta Vf_{m}}+\mathcal{O}\left(e^{-L/L_{0}}e^{-\beta Vf}\right)
\end{equation}
where $f=\min_{m}Re\left[f_{m}\right]$ and $L_{0}$ is of the order
of the largest correlation length of the system. The $f_{m}$'s have
the interpretation of complex free energy densities, and are independent
of $L$. These phases are stable if $Re\left(f_{m}\right)=f$ or metastable
otherwise. The zeros of the partition function are within $\mathcal{O}\left(e^{-L/L_{0}}\right)$
of the solutions of the equations. 
\begin{eqnarray*}
Re\left(f_{m}\right) & = & Re\left(f_{n}\right)=f\\
Im\left(f_{m}\right) & = & Im\left(f_{n}\right)+\left(2p+1\right)\frac{\pi}{\beta V}
\end{eqnarray*}
for some $m\ne n$ and $p\in Z$ . We can apply this directly to region
III, using the representation
\begin{equation}
Z=\sum_{reals}e^{-LE_{r}}+\sum_{pairs}\left(e^{-LE_{p}}+e^{-LE_{p}^{*}}\right)
\end{equation}
of the partition function where we identify $\beta$ with $L$. We
identify $LE_{0}$ and $LE_{0}^{*}$ as $\beta L^{d}f_{0}$ and $\beta L^{d}f_{0}^{*}$,
so that the partition function has a zero for values of the parameters
such that
\begin{equation}
\beta\, Im\left[f_{0}\right]=\frac{\left(2p+1\right)\pi}{2V}
\end{equation}
This tells us that the zeros of the partition function lie on the
boundary of region III, defined by $Im\left[f_{0}\right]=0$, in the
limit $V\rightarrow\infty$. As the volume of the system is taken
to infinity, the zeros of the partition function lie asymptotically
on the boundary between phases. Note that this analysis depends on
$L_{0}$ remaining finite. At a 2nd-order transition, $L_{0}$ goes
to infinity and the approximation is invalid. Zeros of $Z$ in region
III can lead to potentially rapid oscillation of $n$-point functions
, and correlation functions are ill-behaved in the vicinity of such
points.

\section{\label{sec:Classical-statistical-mechanics}$\mathcal{PT}$ symmetry
in classical statistical mechanics }

In this section we consider several models of classical statistical
mechanics that are $\mathcal{PT}$-symmetric. In all of these, it
is convenient to discuss the one-dimensional version of the model,
which is analytically tractable. In addition to models where the classical
Hamiltonian, and hence the transfer matrix, is complex, there are
also models where the classical Hamiltonian and transfer matrix are
real, but the transfer matrix is not symmetric. Such models have a
{}``hidden'' $\mathcal{PT}$ symmetry. In those cases where the
classical Hamiltonian is real, the matrix elements of $T$ are positive.
The Perron-Frobenius theorem applies, and the eigenvalue of $T$ of
greatest magnitude will be real. Such models thus may lie in region
I or II, but never in region III.

\subsection{$Z(N)$ Models}

$Z(N)$ spin systems with complex magnetic fields arise naturally
as simplified models of $SU(N)$ gauge theories non-zero chemical
potential \cite{DeGrand:1983fk}, with the case $N=3$ correspoinding
to QCD at finite baryon density. These models are naturally $\mathcal{PT}$-symmetric.
In these models, there is a clear connection of $\mathcal{PT}$ symmetry
with $Z(N)$ Fourier transforms. This is not surprising: The Fourier
transform of a real function $f(x)$ on $R$ obeys
\begin{equation}
\tilde{f}\left(k\right)^{*}=\tilde{f}\left(-k\right)
\end{equation}
or equivalently
\begin{equation}
\tilde{f}\left(-k\right)^{*}=\tilde{f}\left(k\right)
\end{equation}
which is precisely the statement of $\mathcal{PT}$ symmetry. In spin
models with complex weights based on groups such as $U(1)$, $Z(N)$
and $SU(N)$, $\mathcal{PT}$ symmetry implies that the character
expansion of a $\mathcal{PT}$-symmetric model has real coefficients.
This explains why the flux-tube model \cite{Patel:1983sc,Patel:1983qc}
gives a purely real representation of the same physics of the $Z(3)$
model to which it is dual. $\mathcal{PT}$ symmetry plays a similar
role in the worldline approach to lattice field theories at non-zero
chemical potential \cite{Chandrasekharan:2008gp} . 

On each lattice site $j$ of a $Z(N)$ spin model there is a spin
$w_{j}$, an element of the group $Z(N)$ which may be parametrized
as $w_{j}=\exp\left(2\pi in_{j}/N\right)$ with $n_{j}\in\left\{ 0,1,...,N-1\right\} $
defined modulo $N$ so that $0$ and $N$ are identified. We take
the operator $\mathcal{P}$ to be charge conjugation, acting as $n_{j}\rightarrow-n_{j}$,
or equivalently $w_{j}\rightarrow w_{j}^{*}$. The operator $\mathcal{T}$
is again complex conjugation. Although $\mathcal{P}$ and $\mathcal{T}$
have the same effect on the $w_{j}$'s, one is a linear operator and
the other antilinear. We will show below that $\mathcal{P}$ is implemented
as a unitary matrix in the transfer matrix formalism. The classical
spin-model Hamiltonian $H$ is defined by
\begin{equation}
-\beta\mathcal{H}=\sum_{\left\langle jk\right\rangle }\frac{J}{2}\left(w_{j}w_{k}^{*}+w_{j}^{*}w_{k}\right)+\sum_{j}\left[h_{R}\left(w_{j}+w_{j}^{*}\right)+h_{I}\left(w_{j}-w_{j}^{*}\right)\right]
\end{equation}
 where $\beta=1/T$, $J$, $h_{R}$ and $h_{I}$ are real and the
sum over $\left\langle jk\right\rangle $ represents a sum over nearest-neighbor
pairs. $H$ is trivially $\mathcal{PT}$-symmetric. This class of
models has complex Boltzmann weights for $N\ge3$ when $h_{I}\ne0$.
In the one-dimensional case, it is convenient to write $\mathcal{H}$
in the form

\begin{equation}
-\beta\mathcal{H}=\sum_{j}\left[\frac{J}{2}\left(w_{j}w_{j+1}^{*}+w_{j}^{*}w_{j+1}\right)+\frac{H_{1}}{2}\left(w_{j}+w_{j+1}\right)+\frac{H_{2}}{2}\left(w_{j}^{*}+w_{j+1}^{*}\right)\right]
\end{equation}
 where $H_{1}$ and $H_{2}$ are also real parameters. The partition
function is given by the sum over all spin configurations
\begin{equation}
Z=\sum_{\left\{ w_{j}\right\} }e^{-\beta\mathcal{H}}.
\end{equation}
Associated with the Hamiltonian is a transfer matrix such that $Z=Tr\, T^{N_{s}}$,
where
\begin{equation}
T_{jk}=\exp\left[\frac{J}{2}\left(z^{j}z^{k*}+z^{j*}z^{k}\right)+\frac{H_{1}}{2}\left(z^{j}+z^{k}\right)++\frac{H_{2}}{2}\left(z^{*j}+z^{*k}\right)\right]
\end{equation}
 and $N_{s}$ is the spatial size of the lattice, and $z=\exp\left[2\pi i/N\right]$
is the generator of $Z(N)$. The allowed values of $j$ and $k$ can
be taken to run over either the set $\left\{ 0,1,..,N-1\right\} $
or the set $\left\{ 1,2,..,N\right\} $, and we generally identify
the indices $0$ and $N$. 

We define the paritiy operator $\mathcal{P}$ by $\mathcal{P}_{jk}=\delta_{j,N-k}$,
which satisfies $\mathcal{P}^{2}=1$. Because $z^{N-j}=z^{*j},$ it
is easy to see that
\begin{equation}
\mathcal{P}T\mathcal{P}=T^{*}
\end{equation}
 an equation also satisfied by the Hamiltonian when written in matrix
form. This is the fundmental relation of $\mathcal{PT}$ symmetry,
and can also be written as $\left[\mathcal{PT},\, T\right]=0$.

The discrete Fourier transform, defined by $ $
\begin{equation}
\mathcal{F}_{jk}=\frac{1}{\sqrt{N}}z^{jk},
\end{equation}
 is a symmetric and unitary operator satisfying $\mathcal{F}\mathcal{F}^{+}=I$.
Furthermore, $\mathcal{F}^{2}=\mathcal{F}^{+2}=\mathcal{P}$, so we
have that the Fourier transform of $T$, $\tilde{T}$, obeys
\begin{eqnarray*}
\tilde{T}^{*} & = & \left(FTF^{+}\right)^{*}\\
 & = & F^{*}T^{*}F^{+*}\\
 & = & F^{+}T^{*}F\\
 & = & F^{+}PTPF\\
 & = & F^{+}F^{2}T\left(F^{+}\right)^{2}F\\
 & = & FTF^{+}\\
 & = & \tilde{T}
\end{eqnarray*}
so we see that the Fourier transform of the transfer matrix is indeed
real.

The $\mathcal{PT}$-symmetric spin models have a complex order parameter
coupled to a complex external field. We can prove that $\mathcal{PT}$
symmetry implies that the order parameter is real. Let $w$ be a spin
operator in a $\mathcal{PT}$ -symmetric $Z(N)$ spin system. The
expectation value of $w$ is given by
\begin{equation}
\left\langle w\right\rangle =\frac{Tr\left[wT^{N_{s}}\right]}{Z}
\end{equation}
where $Z=Tr\left[T^{N_{s}}\right]$ and $N_{s}$ is the extent of
the system in the direction in which the transfer matrix $T$ acts.
Using $\mathcal{PT}$ symmetry, we have
\begin{eqnarray*}
\left\langle w\right\rangle  & = & \frac{Tr\left[PwP^{2}T^{N_{s}}P\right]}{Z}\\
 & = & \frac{Tr\left[w^{*}T^{*N_{s}}\right]}{Z}\\
 & = & \left\langle w\right\rangle ^{*}
\end{eqnarray*}
where we have used the fact that $Z$ is real. The reality of $\left\langle w\right\rangle $
is analogous to the result that $\left\langle x\right\rangle $ is
purely imaginary in $\mathcal{PT}$-symmetric quantum mechanics. An
alternative proof of the reality of $\left\langle w\right\rangle $
can be given: in the representation of the system induced by the discrete
Fourier transform, the coefficients of the matrix representations
of both $w$ and $T$ are all real. Equivalently, this can be also
be seen easily from the character expansion of expressions like
\begin{equation}
\exp\left[H_{1}z+H_{2}z^{*}\right]=\sum_{j=0}^{N-1}a_{j}z^{j}
\end{equation}
 where all the coefficients $a_{j}$ are real if $H_{1}$and $H_{2}$
are real.

\begin{figure}
\includegraphics[width=6in]{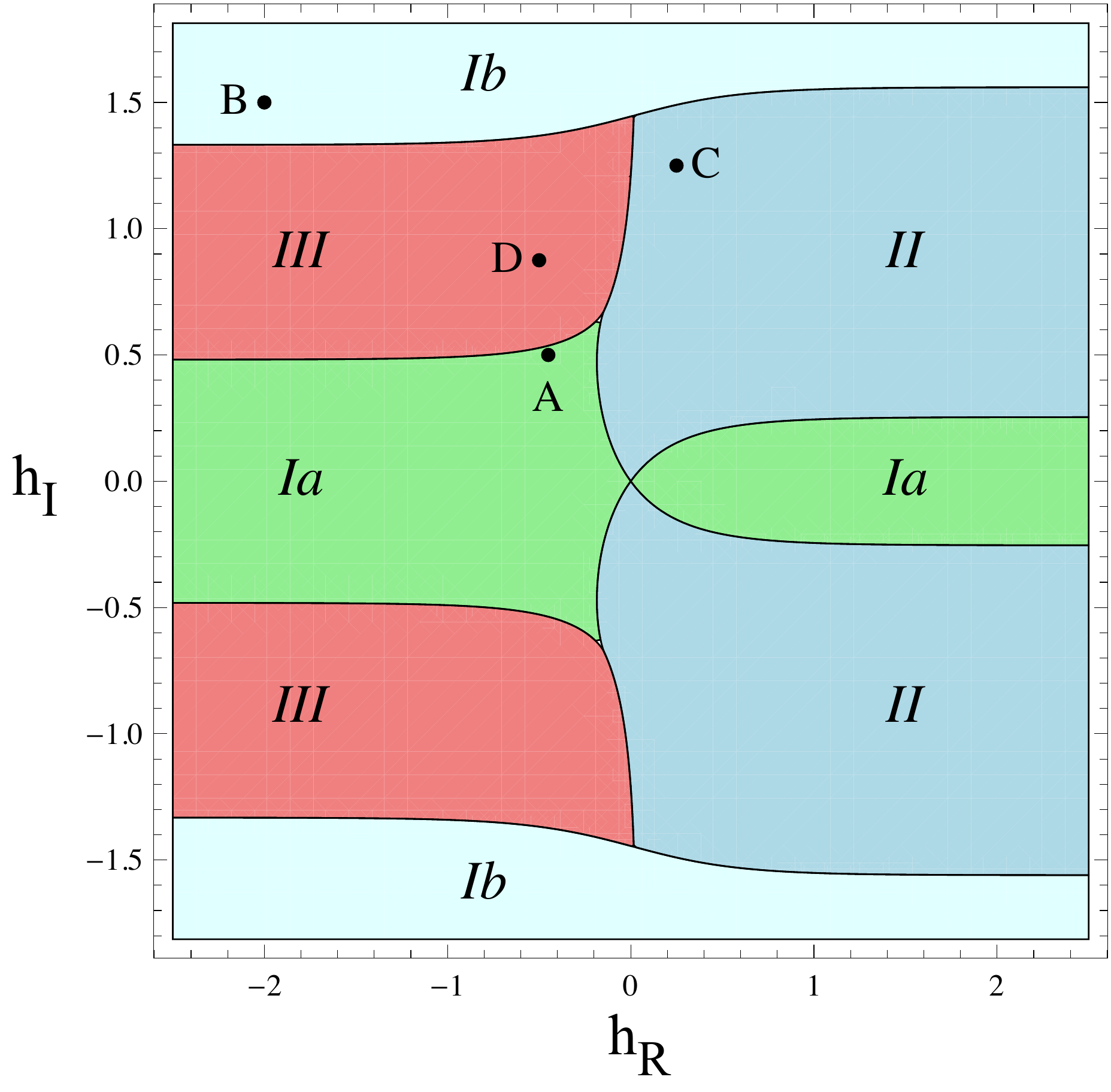}

\caption{\label{fig:Phase-diagram-for-Z3}Phase diagram for the $d=1$ ${\mathcal{P}T}$-symmetric
$Z(3)$ spin model in the $h_{R}-h_{I}$ plane at $J=0.2$. The interpretation
of regions Ia, Ib, II and III are given in the text.}

\end{figure}

\begin{figure}
\includegraphics[width=6in]{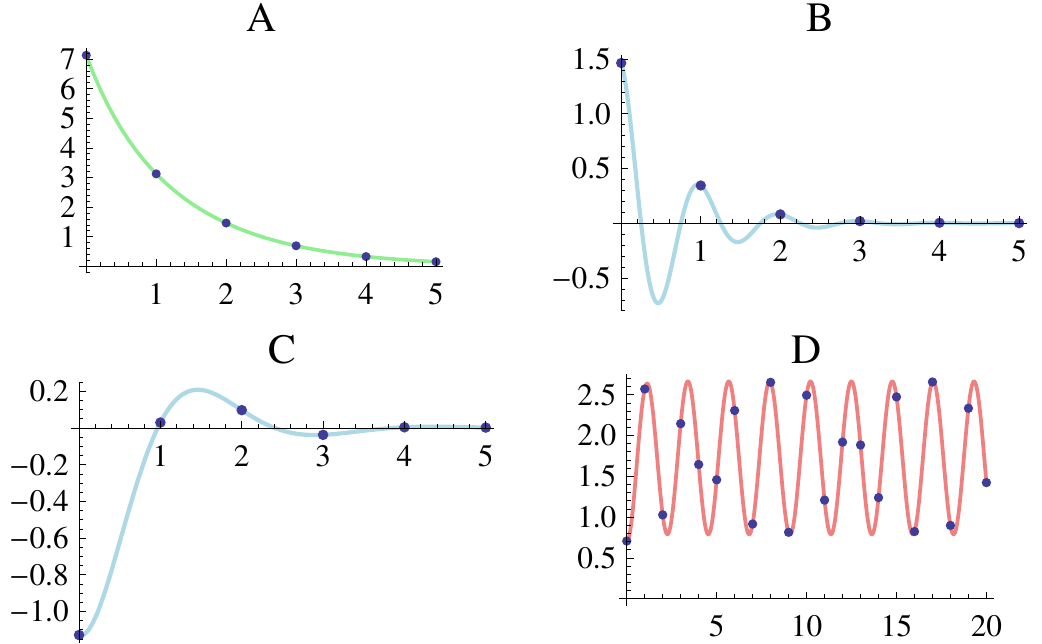}

\caption{\label{fig:The-two-point-function}The two-point function as a function
of lattice spacing for the parameters corresponding to points A, B,
C and D in figure \ref{fig:Phase-diagram-for-Z3}.}

\end{figure}

We illustrate the rich behavior possible in these models using the
case of a $Z(3)$ model in $d=1$ \cite{Ogilvie:2011mw}. If $h_{I}=0$
, then the transfer matrix $T$ is Hermitian. When $h_{I}\neq0$,
$-\beta H$ is no longer real and $T$ is no longer Hermitian, but
is $\mathcal{PT}$ symmetric. Figure \ref{fig:Phase-diagram-for-Z3}
shows the phase diagram in the $h_{R}-h_{I}$ plane for $J=0.2$.
There are four distinct regions. In region Ia, all three eigenvalues
of the transfer matrix are real and positive. This region includes
the line $h_{I}=0$, and has properties similar to those found in
the Hermitian case. In region Ib, all of the eigenvalues are real,
but at least one of them is negative. In region II, the eigenvalue
of $T$ largest in magnitude is real, but the two other eigenvalues
form a complex conjugate pair. In region III, the two eigenvalues
largest in magnitude form a complex conjugate pair, and the third,
smaller, eigenvalue is real. In both region II and region III, $\mathcal{PT}$
symmetry is broken, but in different ways. Borrowing the terminology
from $\mathcal{PT}$-symmetric quantum mechanics, we will describe
the behavior in region III as $\mathcal{PT}$-symmetry breaking of
the ground state, while region II is $\mathcal{PT}$ - symmetry breaking
of an excited state. The behavior of the two-point function $G\left(\left|j-k\right|\right)=\left<w\left(j\right)w^{\dagger}\left(k\right)\right>$
differs substantially in the three regions. In region I, the two-point
function falls off exponentially. We show typical behavior in region
Ia in figure \ref{fig:The-two-point-function} for point A where $\left(h_{R},h_{I}\right)=\left(-0.45,0.5\right)$.
Similar behavior occurs in region Ib, as shown in the figure for point
B where $\left(h_{R},h_{I}\right)=\left(-2.0,1.5\right)$. Although
the figure shows that the continuation of the two-point function away
from integer values can be negative, note that the values at integer
points are all non-negative. The two-point function at point C in
region II where $\left(h_{R},h_{I}\right)=\left(0.25,1.25\right)$
shows the damped oscillatory behavior associated with $\mathcal{PT}$
breaking in excited states. For the point D in region III, where $\left(h_{R},h_{I}\right)=\left(-0.5,0.875\right)$,
the $\mathcal{PT}$ breaking of the ground state leads to oscillatory
behavior of the two-point function in the limit of large distance.
Note that region III only occurs when $h_{R}$ is negative. For $h_{R}<0$
and $h_{I}=0$, the spin configurations with lowest energy have a
two-fold degeneracy. With $h_{I}=0$, the ground state of the transfer
matrix is unique. For the case $h_{R}<0$, $h_{I}=0$, and $J$ large,
the splitting of the two lowest eigenvalues of the transfer matrix
in $d=1$ is small. For sufficiently strong $h_{I}$, the real parts
of the two lowest eigenvalues of $T$ merge, and $\mathcal{PT}$ symmetry
breaking of the ground state occurs.

\subsection{The chiral Potts model}

Out first example of a system with a hidden $\mathcal{PT}$ symmetry
is the chiral Potts model \cite{Ostlund:1981zz,Howes:1983mk}. It
is a variant of $Z(N)$ spin models, and we use the same notation
as above. Consider a $d=1$ $Z(N)$ spin model with Hamiltonian $\mathcal{H}$
of the form
\begin{equation}
-\beta\mathcal{H}=\sum_{j}\left[\frac{J}{2}\left(w_{j}uw_{j+1}^{*}+w_{j}^{*}u^{*}w_{j+1}\right)\right]
\end{equation}
where $u=\exp\left(2\pi i\Delta/N\right)$ with $\Delta\in\left[0,1\right]$
. The classical Hamiltonian $\mathcal{H}$ is real, and the model
is invariant under the action of $\mathcal{T}$, regarded as complex
conjugation. However, the transfer matrix is not Hermitian for general
$\Delta$. The transfer matrix is given by
\begin{equation}
T_{jk}=\exp\left[\frac{J}{2}\left(z^{j}uz^{k*}+z^{j*}u^{*}z^{k}\right)\right]
\end{equation}
where as before $z=\exp\left(2\pi i/N\right)$. The reality of $\mathcal{H}$
implies that $T_{jk}^{*}=T_{jk}$, and thus possesses a generalized
$\mathcal{PT}$ symmetry, with $\mathcal{P}$ simply taken to be the
identity. Although $T$ is real, it is not symmetric in general, because
$T^{T}\ne T$. Because the transfer matrix has only real positive
entries, by the Perron-Frobenius theorem the correlation functions
of the chiral Potts model can exhibit region I or region II behavior,
but never region III.
\begin{equation}
\end{equation}

\subsection{The ANNNI model }

The Anisotropic Next-Nearest-Neighbor Ising (ANNNI) model \cite{Selke:1988ss}
is a prototypical example of a system that appears as if it \textquotedblleft{}should\textquotedblright{}
have a Hermitian transfer matrix, but does not. Instead, the model
has a generalized $\mathcal{PT}$ symmetry which underlies the model's
unusual phase structure \cite{Ogilvie:2011mw}. The one-dimensional
model is exactly solvable, and has a reduced Hamiltonian 
\begin{equation}
\beta H=-K_{1}\sum_{j}s_{j}s_{j+1}-K_{2}\sum_{j}s_{j}s_{j+2}
\end{equation}
 where $K_{1}$ and $K_{2}$ are real couplings and the Ising spins
take on the values $\pm1$. The Hamiltonian is real, and stochastic
simulations of the model may be carried out with ease. One approach
to solving the model is to construct a $4\times4$ transfer matrix
between nearest neighbor pairs 
\begin{equation}
T_{4}=\left(\begin{array}{cccc}
e^{2K_{1}+2K_{2}} & e^{K_{1}} & e^{-K_{1}} & e^{-2K_{2}}\\
e^{-K_{1}} & e^{2K_{2}-2K_{1}} & e^{-2K_{2}} & e^{K_{1}}\\
e^{K_{1}} & e^{-2K_{2}} & e^{2K_{2}-2K_{1}} & e^{-K_{1}}\\
e^{-2K_{2}} & e^{-K_{1}} & e^{K_{1}} & e^{2K_{1}+2K_{2}}
\end{array}\right)
\end{equation}
 The partition function for $N$ spins, with $N$ even and periodic
boundary conditions, is given by $Z=Tr\left[T_{4}^{N/2}\right]$.
The matrix $T_{4}$ is real but not symmetric. It commutes with a
generalized parity operator $\mathcal{P}$ of the form 
\begin{equation}
\mathcal{P}=\left(\begin{array}{cccc}
0 & 0 & 0 & 1\\
0 & 0 & 1 & 0\\
0 & 1 & 0 & 0\\
1 & 0 & 0 & 0
\end{array}\right)
\end{equation}
 which implements the symmetry of the model under $s\rightarrow-s$;
$\mathcal{T}$ is complex conjugation, and acts trivially on $T_{4}$.
The combination of the reality of $T_{4}$ together with $\left[T_{4},P\right]=0$
establishes that $T_{4}$ is $\mathcal{PT}$-symmetric. There is another
approach to solving the one-dimensional model which better displays
its $\mathcal{PT}$ symmetry. We introduce a set of bond variables
$\sigma_{j}$ into the partition function $Z$ which we force to be
equal to $s_{j}s_{j+1}$ via the $Z_{2}$ delta function $\left(1+\sigma_{j}s_{j}s_{j+1}\right)/2$.
We can then write $s_{j}s_{j+2}=\sigma_{j}\sigma_{j+1}$ in the Hamiltonian.
It appears that the new Hamiltonian is simply 
\begin{equation}
\beta H=-K_{1}\sum_{j}\sigma_{j}-K_{2}\sum_{j}\sigma_{j}\sigma_{j+1}
\end{equation}
 and the model reduces to a standard Ising model in an external field.
This is somewhat misleading, because there remains a determinantal
factor associated with the change of variables 
\begin{equation}
Z=\sum_{\left\{ s\right\} }\sum_{\left\{ \sigma\right\} }\prod_{j}\left[\frac{1+\sigma_{j}s_{j}s_{j+1}}{2}\right]e^{-\beta H}
\end{equation}
Carrying out the sum over the s variables with periodic boundary conditions,
we find 
\begin{equation}
Z=\sum_{\left\{ \sigma\right\} }\left[1+\prod_{j}\sigma_{j}\right]e^{-\beta H}
\end{equation}
 which tells us that there is a global constraint on the partition
function: only configurations with $\prod_{j}\sigma_{j}=1$ contribute.
Let $T_{2}$ be the $2\times2$ transfer matrix of the one-dimensional
Ising model in an external field 
\begin{equation}
T_{2}=\left(\begin{array}{cc}
e^{K_{2}+K_{1}} & e^{-K_{2}}\\
e^{-K_{2}} & e^{K_{2}-K_{1}}
\end{array}\right).
\end{equation}
We define another matrix, $\tilde{T}_{2}$, as 
\begin{equation}
\tilde{T}_{2}=\sigma_{3}^{1/2}T_{2}\sigma_{3}^{1/2}=\left(\begin{array}{cc}
1 & 0\\
0 & i
\end{array}\right)\left(\begin{array}{cc}
e^{K_{2}+K_{1}} & e^{-K_{2}}\\
e^{-K_{2}} & e^{K_{2}-K_{1}}
\end{array}\right)\left(\begin{array}{cc}
1 & 0\\
0 & i
\end{array}\right)
\end{equation}
 such that the transfer matrix of the model is the $4\times4$ matrix
$\tilde{T}_{4}=T_{2}\oplus\tilde{T}_{2}$. The square of the eigenvalues
of $\tilde{T}_{4}$ are the eigenvalues of $T_{4}$, as they must
be, and $Z=Tr\left[\tilde{T}_{4}^{N}\right]$. The transfer matrix
$\tilde{T}_{4}$ is invariant under $\mathcal{PT}$, with the parity
operator $\mathcal{P}$ given by $1\oplus\sigma_{3}$, and $\mathcal{T}$
given by complex conjugation. This construction leads directly to
the same eigenvalues found in \cite{Selke:1988ss}. The eigenvalues
of $T_{2}$ are of course always real, while the eigenvalues of $\tilde{T}_{2}$
are either real or form a conjugate pair. For $\cosh K_{1}>e^{-2K_{2}}$,
the eigenvalues of $\tilde{T}_{2}$ are real, and the spin-spin two-point
function decays exponentially. The system is in region I. For $\cosh K_{1}<e^{-2K_{2}}$,
the eigenvalues of $\tilde{T}_{2}$ are complex, and the spin-spin
two-point function shows a periodic modulation of its exponential
decay. This is region II, and the line $\cosh K_{1}=e^{-2K_{2}}$
defines the disorder line separating the two regions. The eigenvalues
of $\tilde{T}_{2}$ are always smaller in absolute value than the
eigenvalues of $T_{2}$ in this model, so region III does not occur
in the $d=1$ ANNNI model.

\section{\label{sec:Quantum-field-theory}$\mathcal{PT}$ symmetry in quantum
statistical mechanics models}

All quantum many-body problems involving a non-zero chemical potential
may be described in terms of a non-Hermitian Hamiltonian with generalized
$\mathcal{PT}$ symmetry \cite{Ogilvie:2011mw}. At first glance,
this is surprising, but it is a simple consequence of the use of Wick
rotation and the Euclidean formalism for equilibrium statistical mechanics.
This $\mathcal{PT}$-symmetric description is closely related to the
sign problem. We will explain in detail how the sign problem arised
in QCD with heavy quarks at non-zero chemical potential. The two-dimensional
case will be solved numerically as an application of $\mathcal{PT}$
symmetry to this class of problems \cite{Ogilvie:2009me,Ogilvie:2011mw}.

\subsection{$\mathcal{PT}$ symmetry at finite density}

We start from a theory with a Hermitian Hamiltonian $H$ and a conserved
global quantum number $N$, obtained from a conserved current $j^{\nu}$,
that commutes with $H$. We assume that $H$ is Hermitian and invariant
under the combined action of time reversal $\mathcal{T}$ and a charge
conjugation $\mathcal{C}$ that reverses the sign of $j^{\nu}$. We
take the number of spatial dimensions to be $d-1$, and the spatial
volume to be $L^{d-1}$. The grand canonical partition function at
temperature $T=\beta^{-1}$ and chemical potential $\mu$ is given
by $Z=Tr\left[\exp\left(-\beta H+\beta\mu N\right)\right]$. If $Z$
is written as a Euclidean path integral, the time component of the
current $j^{0}$ will Wick rotate to $ij^{d}$, while the chemical
potential $\mu$ does not change. This leads directly to a non-positive
weight in the path integral, and is the origin of the sign problem
in finite density calculations. The Euclidean space Lagrangian density
may be written as $\mathcal{L}-i\mu j^{d}$ where $\mathcal{L}$ is
the Euclidean Lagrangian for $\mu=0$; $\mathcal{L}-i\mu j^{d}$ is
complex. The nature of the problem is changed by changing the direction
of Euclidean time, so that we are now considering a problem at zero
temperature with one compact spatial dimension of circumference $\beta$.
Upon returning to Minkowski space, \textmu{} does not rotate. We pick,
say, the $\nu=1$ direction to be the new time direction and the new
inverse temperature is $L$. When $\mu=0$, the original Hamiltonian
is obtained, but for $\mu\ne0$ the partition function is now given
by
\begin{equation}
Z=Tr\left[e^{-LH_{\beta}}\right]
\end{equation}
 where
\begin{equation}
H_{\beta}=H-i\mu\int d^{d-1}x\, j^{_{d}}.
\end{equation}
 The new Hamiltonian $H_{\beta}$ is non-Hermitian, but possesses
a generalized $\mathcal{PT}$ symmetry, where the role of $\mathcal{P}$
is played by the charge conjugation operator $\mathcal{C}$ that changes
the sign of $j^{0}$ and $N$. Under the combined action of $\mathcal{CT}$,
$j^{d}\rightarrow-j^{d}$ and $i\rightarrow-i$, leaving the Hamiltonian
$H_{PT}$ invariant. If we introduce the operator $H_{L}=H-\mu N$,
we have the relation
\begin{equation}
Z=Tr\left[e^{-\beta H_{L}}\right]=Tr\left[e^{-LH_{\beta}}\right]
\end{equation}
induced by the space-time transformation that exchanges directions
$1$ and $d$. Note that $Z$ is obtained from $H_{L}$ by a sum over
all eigenstates, but is dominated by the ground state of $H_{\beta}$
in the limit of large $L$.

\subsection{$d=2$ gauge theories}

Within the Euclidean space formalism, a non-zero temperature $T$
is obtained by making the bosonic fields periodic in Euclidean time,
with period $\beta=1/T$. On the other hand, a non-zero chemical potential
must be implemented in a way that makes the weight function used in
the Feynman path integral complex, as we have seen above. We will
show below exactly how QCD with quarks at finite density may be interpreted
as a theory with $\mathcal{PT}$ symmetry. 

\begin{figure}
\includegraphics[width=6in]{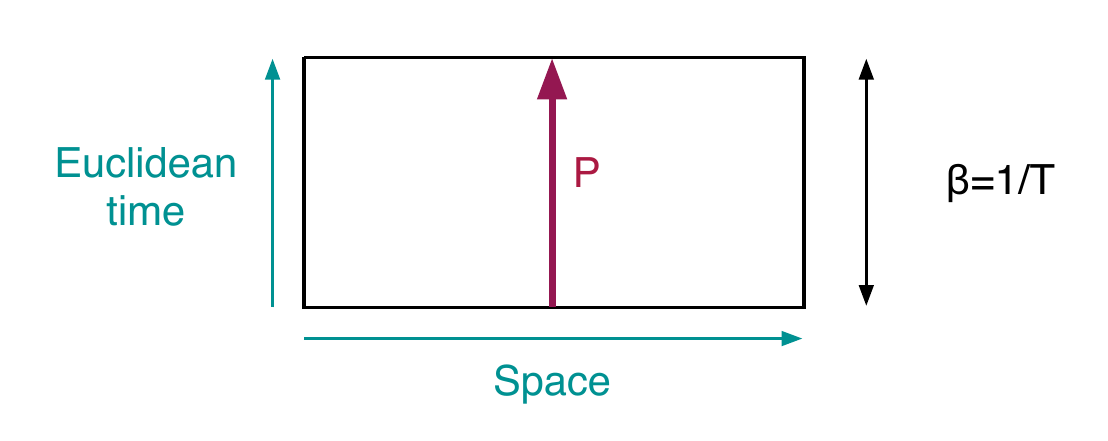}

\caption{\label{fig:The-Polyakov-loop}The Polyakov loop in Euclidean space-time.}

\end{figure}

The Polyakov loop plays a crucial role. Defined as a path-ordered
exponential of the gauge field, in $3+1$ dimensions the Polyakov
loop operator $P$ is given by 
\begin{equation}
P\left(\vec{x}\right)=\mathbb{P}\exp\left[i\int_{0}^{\beta}dtA_{4}\left(\vec{x},t\right)\right],
\end{equation}
and represents the insertion of a static quark into a thermal system
of gauge fields at a temperature $T=\beta^{-1}$. Figure \ref{fig:The-Polyakov-loop}
shows the Polyakov loop in this geometry. Because of the periodic
boundary conditions in the Euclidean time direction, the Polyakov
loop is a closed loop, and its trace is gauge invariant. Also known
as the Wilson line, the Polyakov loop represents the insertion of
a static quark at a spatial point $\vec{x}$ in a gauge theory at
finite temperature. In particular, the thermal average of the trace
of $P$ in an irreducible representation $R$ of the gauge group is
associated with the additional free energy $F_{R}$ required to insert
a static quark in the fundamental representation via 
\begin{equation}
\left\langle Tr_{R}P\left(\vec{x}\right)\right\rangle =e^{-\beta F_{R}}.
\end{equation}
Pure $SU(N)$ gauge theories have a global $Z(N)$ symmetry $P\rightarrow zP$
where $z=e^{\frac{2\pi i}{N}}$ is the generator of $Z(N)$, the center
of $SU(N)$. This symmetry, if unbroken, guarantees that for the fundamental
representation $F$, $\left\langle Tr_{F}P\left(\vec{x}\right)\right\rangle =0$.
This is interpreted as $F_{F}$ being infinite, and an infinite free
energy is required to insert a heavy quark into the system. On the
other hand, if the $Z(N)$ symmetry is spontaneously broken, the free
energy required is finite. Thus confinement in pure gauge theories
is associated with unbroken center symmetry, and broken symmetry with
a deconfined phase. The Polyakov loop is the order parameter for the
deconfinement transition in pure gauge theories $\left<Tr_{F}P\right>=0$
in the confined phase and $\left<Tr_{F}P\right>\ne0$ in the deconfined
phase. The addition of dynamical quarks in the fundamental representation
explicitly breaks this $Z(N)$ symmetry. Nevertheless, the Polyakov
loop remains important in describing the behavior of the system, as
we will see in our treatment of the sign problem.

In pure gauge theories, the Wilson loop operator is used to measure
the string tension between quarks in the confined phase where $F_{R}$
vanishes for representations transforming non-trivially under $Z(N)$.
At non-zero temperature, a timelike string tension $\sigma_{k}^{(t)}$
between k quarks and k antiquarks can be measured from the behavior
of the correlation function 
\begin{equation}
\left<Tr_{F}P^{k}\left(\vec{x}\right)Tr_{F}\left(P^{+}\left(\vec{y}\right)\right)^{k}\right>\simeq\exp\left[-\frac{\sigma_{k}^{(t)}}{T}\left|\vec{x}-\vec{y}\right|\right]
\end{equation}
at sufficiently large distances. A confining phase is defined by two
properties: the expectation value $\left<Tr_{R}P\right>$ is zero
for all representations $R$ transforming non-trivially under $Z(N)$,
and the string tensions $\sigma_{k}^{(t)}$ must be non-zero for $k=1$
to $N-1$. 

Perturbation theory can be used to calculate the one-loop free energy
density $f_{q}$ of quarks in $d+1$ dimensions in the fundamental
representation with spin degeneracy $s$ moving in a Polyakov loop
background at non-zero temperature $T=\beta^{-1}$ and chemical potential
$\mu$ 
\begin{equation}
f_{q}=-sT\int\frac{d^{d}k}{\left(2\pi\right)^{d}}Tr_{R}\left[\ln\left(1+Pe^{\beta\mu-\beta\omega_{k}}\right)+\ln\left(1+P^{+}e^{-\beta\mu-\beta\omega_{k}}\right)\right]
\end{equation}
 where $\omega_{k}=\sqrt{k^{2}+M^{2}}$ is the energy of the particle
as a function of $k$ and $M$ is the mass of the particle \cite{Gross:1980br,Weiss:1980rj}.
The expression for a bosonic field is similar. The logarithm can be
expanded to give
\begin{equation}
f_{q}=sT\int\frac{d^{d}k}{\left(2\pi\right)^{d}}\sum_{n=1}^{\infty}\frac{\left(-1\right)^{n}}{n}\left[e^{n\beta\mu-n\beta\omega_{k}}Tr_{R}P^{n}+e^{-n\beta\mu-n\beta\omega_{k}}Tr_{R}P^{+n}\right].
\end{equation}
 This expresssion has a simple interpretation as a sum of paths winding
around the timelike direction. With standard boundary conditions,
which are periodic for bosons and antiperiodic for fermions, this
one-loop free energy always favors the deconfined phase.

The effects of heavy quarks in the fundamental representation, with
$\beta M\gg1$, on the gauge theory can be obtained approximately
from the $n=1$ term in the free energy
\begin{equation}
f_{q}\approx-sT\int\frac{d^{d}k}{\left(2\pi\right)^{d}}Tr_{F}\left[Pe^{\beta\mu-\beta\omega_{k}}+P^{+}e^{-\beta\mu-\beta\omega_{k}}\right]
\end{equation}
 because term with higher $n$ are suppressed by a factor $e^{-n\beta M}$.
In this approximation, bosons and fermions have the same effect at
leading order. After integrating over $k$, the free energy $f_{q}$
can be written as $f_{q}\approx-h_{F}\left[e^{\beta\mu}Tr_{F}P+e^{-\beta\mu}Tr_{F}P^{+}\right]$.
The one-loop free energy density is the one-loop effective potential
at finite temperature. Thus the free energy for the heavy quarks can
be added to the usual gauge action to give an effective action which
involves only the gauge fields. The effective action is given by 
\begin{equation}
S_{eff}=\int d^{d+1}x\left[\frac{1}{4g^{2}}\left(F_{\mu\nu}^{a}\right)^{2}-h_{F}\left(e^{\beta\mu}Tr_{F}(P)+e^{-\beta\mu}Tr_{F}(P^{+})\right)\right]
\end{equation}
and the structure and symmetries of the theory are obviously the same
in any number of spatial dimensions. Because $Tr_{F}P$ is complex
for $N\ge3$, the effective action for the gauge fields is complex.
This is a form of the so-called sign problem for gauge theories at
finite density: the Euclidean path integral involve complex weights.
This problem is a fundamental barrier to lattice simulations of QCD
at finite density.

\subsection{Heavy quarks at $\mu\ne0$ in two dimensions and $\mathcal{PT}$symmetry}

\begin{figure}
\includegraphics{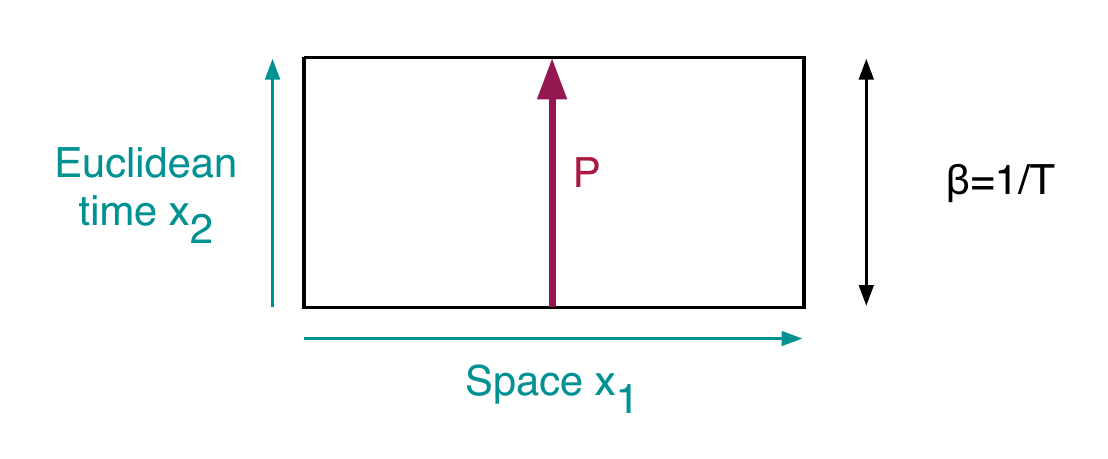}

\caption{\label{fig:Polyakov-loop-2d}The Polyakov loop in $(1+1)$-dimensional
space-time. }

\end{figure}

\begin{figure}
\includegraphics[width=2.5in]{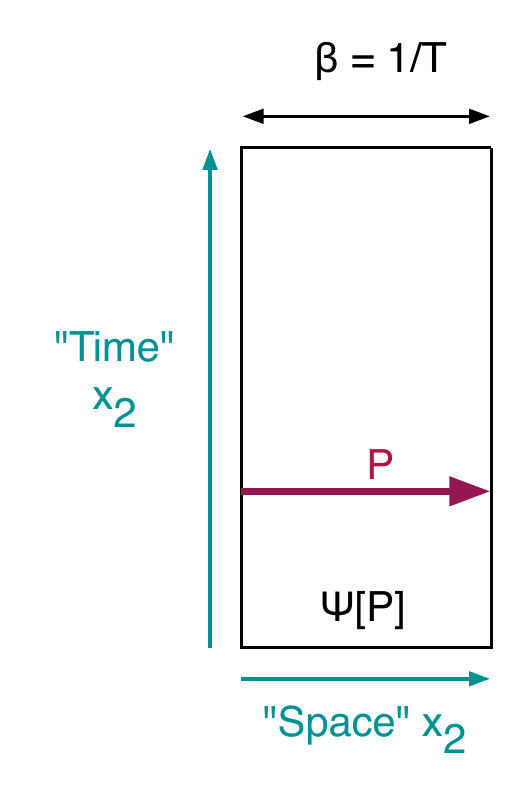}

\caption{\label{fig:The-Polyakov-loop-transfer}The Polyakov loop in a $(1+1)$-dimensional
transfer matrix geometry.}

\end{figure}

In one space and one time dimension, the field theory arising from
the effective action $S_{eff}$ can be reduced to a $\mathcal{PT}$-symmetric
Hamiltonian acting on class functions of the gauge group. The effective
action, including the effects of heavy quarks, is 
\begin{equation}
S_{eff}=\int d^{2}x\left[\frac{1}{4g^{2}}\left(F_{\mu\nu}^{a}\right)^{2}-h_{F}\left(e^{\beta\mu}Tr_{F}(P)+e^{-\beta\mu}Tr_{F}(P^{+})\right)\right]
\end{equation}
where the gauge field $A_{\mu}$ now has two components. Figure \ref{fig:Polyakov-loop-2d}
shows the Polyakov loop in a $1+1$-dimensional geometry. It is convenient
to work in a gauge where $A_{1}=0$; this is turn implies that $A_{2}$
depends only on $x_{1}$. After integration over $x_{2}$, we are
left with a Lagrangian 
\begin{equation}
L=\frac{\beta}{2g^{2}}\left(\frac{dA_{2}^{a}}{dx_{1}}\right)^{2}-h_{F}\beta\left[e^{\beta\mu}Tr_{F}(P)+e^{-\beta\mu}Tr_{F}(P^{+})\right]
\end{equation}
which we regard as the Lagrangian for a system evolving as a function
of a time coordinate $x_{1}$. This represents a change from a Euclidean
time point of view to a transfer matrix geometry, as shown in figure
\ref{fig:The-Polyakov-loop-transfer}. In this geometry, the Polyakov
loop represents the insertion of an electric flux line in a box with
periodic boundary conditions, and the free energy density is obtained
from the lowest-lying eigenvalue of the transfer matrix.

The physical states of the system are gauge-invariant, meaning that
they are class functions of $P$: $\Psi\left[P\right]=\Psi\left[gPg^{+}\right]$.
The group characters form an orthonormal basis on the physical Hilbert
space: $\Psi\left[P\right]=\sum_{R}a_{R}Tr_{R}\left(P\right)$. The
Hamiltonian $H$, obtained from $L$, acts on the physical states
as 
\begin{equation}
H=\frac{g^{2}\beta}{2}C_{2}-h_{F}\beta\left[e^{\beta\mu}Tr_{F}(P)+e^{-\beta\mu}Tr_{F}(P^{+})\right]
\end{equation}
 where $C_{2}$ is the quadratic Casimir operator for the gauge group,
the Laplace-Beltrami operator on the group manifold. We have thus
reduced the problem of heavy quarks at finite density in two dimensions
to a problem of quantum mechanics on the gauge group. Unforturnately,
the Hamiltonian $H$ is not Hermitian when $\mu\ne0$, and thus cannot
be relied upon to have real eigenvalues. This is a direct manifestation
of the sign problem.

Although the Hamiltonian $H$ is not Hermitian when $\mu\neq0$, it
is $\mathcal{PT}$-symmetric under the transformations 
\begin{equation}
\mathcal{P}:x_{2}\rightarrow-x_{2}\,\,\, A_{2}\rightarrow-A_{2}
\end{equation}
 
\begin{equation}
\mathcal{T}:i\rightarrow-i
\end{equation}
which should be regarded as parity and time-reflection in the transfer
matrix geometry. Together these lead to 
\begin{equation}
\mathcal{PT}:P\rightarrow P
\end{equation}
which leaves the Hamiltonian invariant. If this $\mathcal{PT}$-symmetry
is unbroken, the eigenvalues of the Hamiltonian will be real, and
there is no sign problem. The $\mathcal{PT}$ symmetry remains even
in the high-density limit where the quark mass $M$ and chemical potential
$\mu$ are taken to infinity in such a way that antiparticles are
suppressed and $P^{+}$ does not appear in $H$. 

The simplest non-trivial gauge group is $SU(3)$, because the cases
of $U(1)$ and $SU(2)$ are atypical. For the gauge group $U(1)$,
the Hamiltonian $H$ may be written as 
\begin{equation}
H=-\frac{e^{2}\beta}{2}\frac{d^{2}}{d\theta^{2}}-h_{F}\beta\left(e^{\beta\mu+i\theta}+e^{-\beta\mu-i\theta}\right)
\end{equation}
 but a simple change of variable $\theta\rightarrow\theta+i\beta\mu$
eliminates $\mu$:
\begin{equation}
H=-\frac{e^{2}\beta}{2}\frac{d^{2}}{d\theta^{2}}-h_{F}\beta\left(e^{+i\theta}+e^{-i\theta}\right)
\end{equation}
This is very similar to the case of the two-dimensional $\mathcal{PT}$-symmetric
sine-Gordon model considered in \cite{Bender:2005hf}. In the case
of $SU(2)$, all the irreducible representations are real, and the
Hamiltonian is Hermitian:
\begin{equation}
H_{SU(2)}=\frac{g^{2}\beta}{2}C_{2}-2h_{F}\cosh\left(\beta\mu\right)\chi_{j=1/2}(P).
\end{equation}
In general there is no sign problem in $SU(2)$ gauge theories at
finite density holds in general, and this feature has been exploited
in lattice simulations with $\mu\ne0$ \cite{Hands:1999md,Kogut:2001na}.

\begin{figure}
\includegraphics[width=4in]{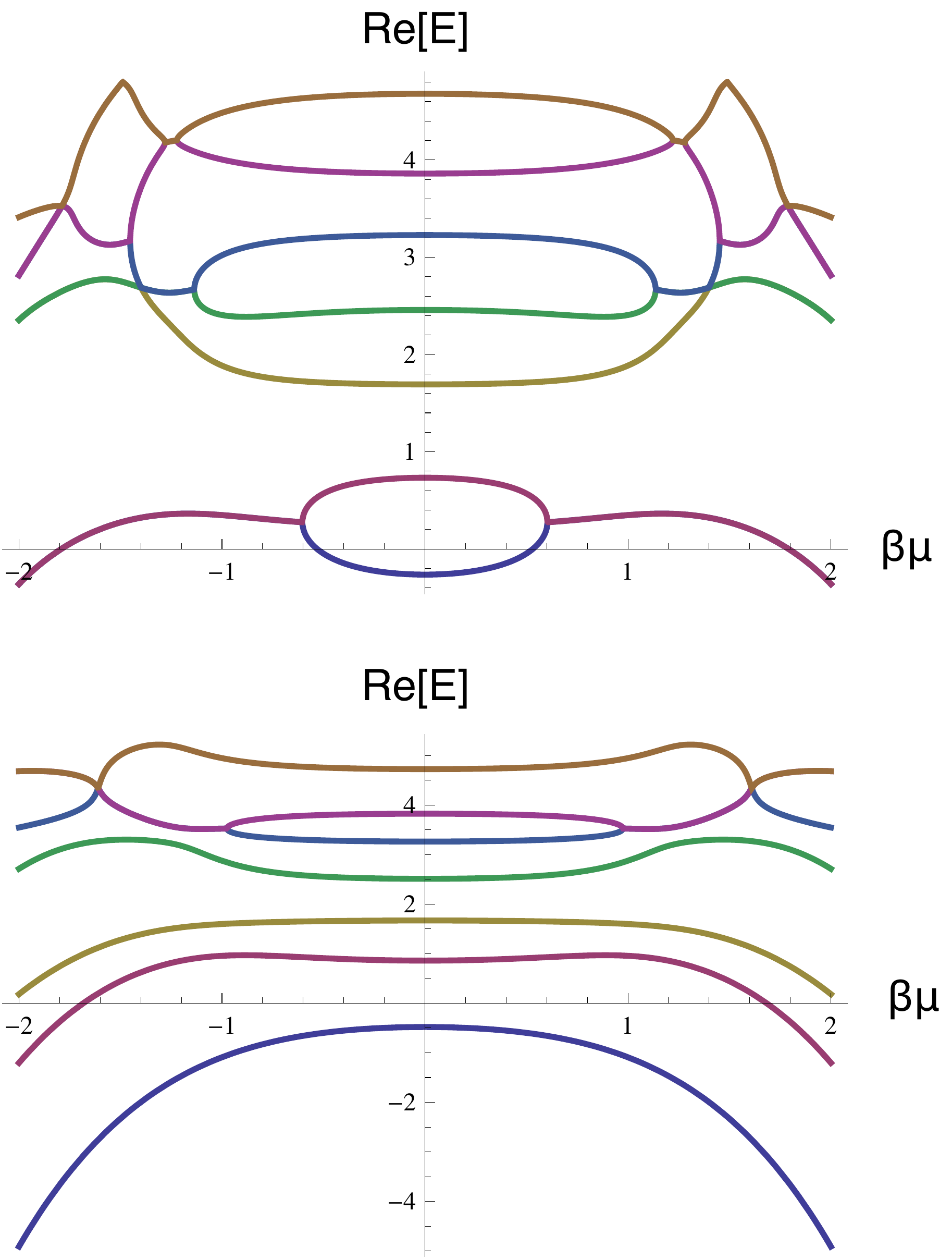}

\caption{\label{fig:su3-both}The real part of the $SU(3)$ Hamiltonian $H_{\beta}$
as a function of $\beta\mu$. The upper graph is for periodic boundary
conditions for the heavy quarks, while the lower graph is for antiperiodic
boundary conditions. The energy has been scaled such that $g^{2}\beta/2$
is set equal to $1$. }
\end{figure}

Thus $N=3$ is the first non-trivial case for $SU(N)$ gauge groups.
We have calculated the lowest eigenvalues of $H$ using finite dimensional
approximants. It is convenient to work in the group character basis.
The Casimir operator $C_{2}$ is diagonal in this basis, and characters
act as raising and lowering operators. For example, in the $4\times4$
subspace spanned by the $1$, $3$ , $\bar{3}$, and $8$ representations
of $SU(3)$, the Hamiltonian takes the form
\begin{equation}
\left(\begin{array}{llll}
0 & e^{-\beta\mu}h_{F}\beta & e^{\beta\mu}h_{F}\beta & 0\\
e^{\beta\mu}h_{F}\beta & \frac{4}{3}\cdot\frac{g^{2}\beta}{2} & e^{-\beta\mu}h_{F}\beta & e^{\beta\mu}h_{F}\beta\\
e^{-\beta\mu}h_{F}\beta & e^{\beta\mu}h_{F}\beta & \frac{4}{3}\cdot\frac{g^{2}\beta}{2} & e^{-\beta\mu}h_{F}\beta\\
0 & e^{-\beta\mu}h_{F}\beta & e^{\beta\mu}h_{F}\beta & 3\cdot\frac{g^{2}\beta}{2}
\end{array}\right).
\end{equation}
 If $h_{F}$ is set to zero, we see that the eigenvalues are proportional
to Casimir invariants $0$, $4/3$ , $4/3$, and $3$ for the $1$,
$3$, $\bar{3}$, and $8$ representations of $SU(3)$. We have therefore
removed an overall factor of $g^{2}\beta/2$, so the overall strength
of the potential term is controlled by the dimensionless parameter
$2h_{F}/g^{2}$. The resulting dimensionless energy eigenvalues are
thus normalized to give the quadratic Casimir operator when $2h_{F}/g^{2}=0$.
The lowest eigenvalues have been calculated numerically using a basis
of dimension nine or larger, with the stability of the lowest eigenvalues
checked by changing the basis size.

The parameter $h_{F}$ is positive for fermions with antiperiodic
boundary conditions in the timelike direction, which are required
for spectral positivity. However, it is also of interest to consider
the case of periodic boundary conditions for the heavy quarks, corresponding
to $h_{F}<0$ \cite{Myers:2007vc,Unsal:2008ch,Myers:2009df}. In figure
\ref{fig:su3-both}, we show the real part of the eigenvalues of $H_{\beta}$,
measured in units where $g^{2}\beta/2$ is set to $1$. The overall
strength of the potential term is set by the dimensionless parameter
$2h_{F}/g^{2}$. In the upper graph, $2h_{F}/g^{2}=-0.5$, corresponding
to periodic boundary conditions for the heavy quarks. The lower graph
shows the real part of the energy eigenvalues for $2h_{F}/g^{2}=0.5$.
In both cases, we see the real parts of pairs of energy eigenvalues
coalescing as $\beta\mu$ is increased. At the point where the real
parts become identical, these energy eigenvalues acquire an imaginary
part, indicative of broken $\mathcal{PT}$ symmetry. In the case of
periodic boundary conditions, we see that the ground state shows $\mathcal{PT}$-symmetry
breaking before any of the higher states; thus for large $\beta\mu$
this places the system in region III. Note that for $N\ge3$, the
heavy quark finite density problem of $SU(N)$ gauge theory is in
the universality class of the Lee-Yang problem for $Z(N)$ spin systems
\cite{DeGrand:1983fk}. In the physical case of antiperiodic boundary
conditions, $\mathcal{PT}$-symmetry breaking appears to occur only
in excited states. In the case where all eigenvalues are real, which
appears to hold for small $\beta\mu$, $\mathcal{PT}$symmetry is
unbroken and the system is in region I. This in turn implies that
for small $\beta\mu$ the sign problem may be solved in principle
by a similarity transform to a Hermitian Hamiltonian. For large $\beta\mu$,
the $\mathcal{PT}$ symmetry is broken in some of the excited states,
which will lead to the region II behavior of sinusoidal decay of spatial
correlation functions at high density.

\section{Conclusions}

For Hermitian systems, there is a well-developed understanding of
critical behavior and phase structure, connecting a wide range of
systems from simple classical spin systems to exotic quantum field
theories. For $\mathcal{PT}$-symmetric models, we are in a sense
starting over again with a richer, larger class of systems. There
are indications that $\mathcal{PT}$ symmetry is crucial to the understanding
of the sign problem. More generally, the characterization of the phase
structure and universality classes of $\mathcal{PT}$-symmetric systems
is a logical extension of the successful effort to characterize critical
phenomena in Hermitian systems.

\acknowledgements{Michael Ogilvie wishes to acknowledge the support of his research by
the US Department of Energy.}

\bibliographystyle{plain}
\bibliography{PT_Review}

\end{document}